\documentclass[letter]{emulateapj}
\usepackage{natbib}
\usepackage{epsfig}
\usepackage{amssymb}
\usepackage{latexsym}
\usepackage{ifthen}
\usepackage{url}
\usepackage{longtable}

\newcommand{\msun}{\,M_{\odot}}

\newcommand{\fdeg}{^{\circ}}

\begin{document}

\title[Young pulsars in GCs]{Young Radio Pulsars in Galactic Globular Clusters}


\author{J.~Boyles$^*$, D.~R.~Lorimer$^*$, P.~J.~Turk$^{\dag}$, R.~Mnatsakanov$^{\dag}$,
R.~S.~Lynch$^{**}$, S.~M.~Ransom$^{\ddag}$, P.~C.~Freire$^{\S}$, and K.~Belczynski$^{\P}$
}
\altaffiltext{$*$}{Department of Physics, 210 Hodges Hall, West Virginia University,
Morgantown, WV 26506}
\altaffiltext{$\dag$}{Department of Statistics, 423 Hodges Hall, West Virginia University,
Morgantown, WV 26506}
\altaffiltext{**}{Department of Astronomy, University of Virginia, P.O. Box 400325, Charlottesville, VA 22904}
\altaffiltext{$\ddag$}{National Radio Astronomy Observatory, 520 Edgemont Rd, Charlottesville, VA}
\altaffiltext{$\S$}{Max-Planck-Institut f$\ddot{u}$r Radioastronomie, Auf dem H$\ddot{u}$gel D-53121, Bonn, Germany}
\altaffiltext{$\P$}{Astronomical Observatory, University of Warsaw, AL Ujazdowskie 4,00-478, Warsaw, Poland} 

\begin{abstract}

Currently three isolated radio pulsars and one binary radio pulsar
with no evidence of any previous recycling are known in 97 surveyed
Galactic globular clusters.  As pointed out by Lyne et al., the
presence of these pulsars cannot be explained by core-collapse
supernovae, as is commonly assumed for their counterparts in the
Galactic disk.  We apply a Bayesian analysis to the results from
surveys for radio pulsars in globular clusters and find the number of
potentially observable non-recycled radio pulsars present in all
clusters to be $<3600$. Accounting for beaming and retention
considerations, the implied birth rate for any formation scenario for
all 97 clusters is $<0.25$ pulsars per century assuming a Maxwellian
distribution of velocities with a dispersion of 10~km~s$^{-1}$.  The
implied birth rates for higher velocity dispersions are substantially
higher than inferred for such pulsars in the Galactic disk.  This
suggests that the velocity dispersion of young pulsars in globular
clusters is significantly lower than those of disk pulsars.  These
numbers may be substantial overestimates due to the fact that the
currently known sample of young pulsars is observed only in metal-rich
clusters. We propose that young pulsars may only be formed in globular
clusters with metallicities with log[Fe/H]~$>-0.6$.  In this case,
the potentially observable population of such young pulsars is
$447^{+1420}_{-399}$ (the error bars give the $95\%$ confidence
interval) and their birth rate is $0.012^{+0.037}_{-0.010}$ pulsars
per century.  The mostly likely creation scenario to explain these
pulsars is the electron capture supernova of a OMgNe white dwarf.
\end{abstract}


\maketitle

\section{Introduction}\label{sec:intro}\setcounter{footnote}{0} 

Since the first discovery of PSR~B1821$-$24 in M28 (Lyne et al. 1987),
143 pulsars\footnote{For an up-to-date list of known globular cluster pulsars,
see http://www.naic.edu/$\sim$pfreire/GCpsr.html}
have been discovered in 27 globular clusters (GCs). While the majority
of GC pulsars are thought to have been formed in low-mass X-ray binary
systems, in which the neutron star has been spun up to millisecond
periods by the transfer of matter from an evolved companion (see, e.g.
Camilo \& Rasio 2005 for a recent review), a small minority of the
GC pulsar population appear to be similar to the isolated ``normal'' pulsars
which inhabit the disk of our Galaxy (i.e. spin periods ($P$) of several hundred
ms, characteristic ages ($\tau_c$) of $10^{7}$--$10^{8}$ yr,
and inferred dipole magnetic field strengths ($B$) of $10^{11}$--$10^{12}$~Gauss).
These pulsars were originally discussed by Lyne et al.~(1996) who noted
that, since no significant star formation has occurred in GCs in the last
billion years (Briley et al. 1994), this population appears to be highly anomalous.  
Lyne et al.~(1996) also note that the young pulsars appear in metal-rich 
GCs.  This trend has persisted despite fifteen years of intense searches 
of most of the cluster population.  A goal of this work is to investigate the 
statistical and astrophysical significance of this result.  

Recent improvements in observational systems over the past decade have
led to a wealth of discoveries of pulsars in GCs.
Sensitive radio surveys for pulsars have been conducted on almost 100
GCs to date. The two most fruitful have been Terzan~5 (Ransom et al. 2005)
and 47~Tucanae (Freire et al.~2001) with 34 and 23 pulsars respectively.
Neither of these clusters harbor any normal pulsars, which we will henceforth
define as being pulsars with spin periods $P>100$~ms and inferred
magnetic field strengths $B>10^{11}$~Gauss. This is somewhat surprising,
given that the significant selection effects 
known to hamper the detection of binary and millisecond pulsars
in clusters (Camilo \& Rasio 2005) are not as severe for normal
isolated pulsars.  

Clearly, some physical mechanism is at work which produces these
apparently young pulsars in GCs in a different way to how we believe
they are formed in the disk of our Galaxy.  One possible method
is the collapse of a white dwarf via an electron capture supernova 
in a binary or collisional system (Ivanova et al. 2008).
In this paper, we revisit the statistics of the normal pulsars
based on the results of recent surveys, and recent studies of the 
Galactic population of normal pulsars (Faucher-Gigu\'ere \& Kaspi 2006
(hereafter FK06);
Ridley \& Lorimer 2010). In \S \ref{sec:sample}, we review the
current sample of normal pulsars in GCs. In \S \ref{sec:fluxlimits}, we
compile a list of flux-density detection limits for 97 GCs based on published
searches and some recent unpublished results. In \S \ref{sec:statistics}, 
we use these limits to characterize the population of young pulsars. 
In \S \ref{sec:intrinsic} we discuss the intrinsic population and birth
rate of young pulsars in GCs. In \S \ref{sec:discussion} we discuss 
formation scenarios and future work with young globular cluster pulsars.

\section{The current sample of young pulsars in GCs}\label{sec:sample} 

\begin{figure}[t]
\includegraphics[angle=270,scale=0.45]{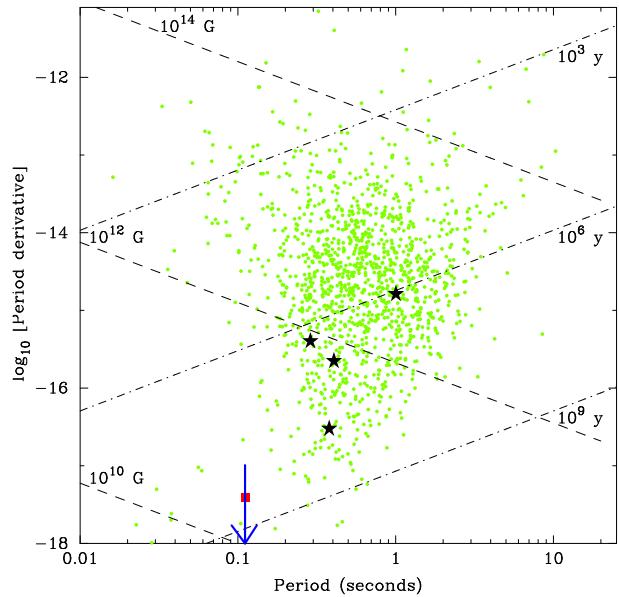}
\caption{\label{fig:ppdot}
$P$--$\dot{P}$ diagram showing young globular cluster pulsars
as black stars and PSR~J1750$-$37A as a red square.  
PSR~B2127+11A is shown as a blue arrow with the arrow representing
the limit on period derivative at the bottom of the diagram.
Dashed lines are lines of
constant magnetic field and dot dash lines are lines of 
constant characteristic age.   
The pulsar population data are taken from the ATNF pulsar database
(Manchester et al 2005).}  
\end{figure}

At least three, possibly four, young pulsars are known in two or three GCs. 
These four pulsars are B1718$-19$ in NGC~6342, B1745$-20$ in NGC~6640, and
J1823$-3021$B and J1823$-3021$C in NGC~6624.  The properties of these pulsars 
are summarized in Table~\ref{tab:sample} and their GC properties
relevant to this data analysis can be seen in Table~\ref{table:gclim}. As can be seen 
by inspecting their position with respect to other normal Galactic field pulsars on the 
$P$--$\dot{P}$ diagram shown in Figure~\ref{fig:ppdot}, these pulsars
appear to be consistent with the distribution of normal pulsars in the
Galactic disk. 

Two long-period pulsars are not included in this compilation. Firstly,
the 110-ms pulsar B2127+11A in GC NGC~7078 which is known to have
a negative period derivative due to contamination by the cluster
potential and nearby stars (Wolszczan et al. 1989). As discussed by these authors,
the pulsar most likely has a small intrinsic period derivative indicating
that it has undergone some recycling in a binary system which has
subsequently been disrupted due to close encounters in the cluster.
It is also noted that a recent measurement of the second period derivative 
of $\ddot{P}$ = $3.2 \times 10^{-29}$ $s^{-1}$ for this pulsar is
entirely consistent with encounters with nearby
stars in the cluster (Jacoby et al.~2006). Another long-period pulsar
which we do not consider to be young is the 110-ms pulsar
J1750$-$37A in GC NGC~6441 (Freire et al. 2008). This is a member of an eccentric
binary system and its low period derivative places it closer to the 
region of the $P$--$\dot{P}$ diagram occupied by the double neutron star binary systems
and the eccentric neutron star--white dwarf binaries (see Figure~\ref{fig:ppdot}).

One other caveat needs to be discussed with the sample of young pulsars present in 
this paper.  PSR~B1718$-$19 may not truly be a member of NGC~6342 and is the only
pulsar located near NGC~6342.  An earlier 
discussion of this subject is presented in Bailes et al. (2005) and a further 
expansion will be presented here.  The original paper used PSR~B1718$-$19's position,
binary status, and dispersion measure (DM) to argue for its association with NGC~6342. 
Its position of 2.3\arcmin \ away from the cluster's center is 3 times greater then the 
half-mass radius of the cluster.  All other young pulsars presented in 
Table~\ref{tab:sample} and a majority of all globular cluster pulsars with timing 
solutions are found within the half-mass radius.  The DM of 75.7~pc~cm$^{-3}$ for 
PSR~B1718$-$19 is somewhat below the predicted values of 120~pc $\rm cm^{-3}$ using 
DM = 20/sin($b$) pc $\rm cm^{-3}$ (Lyne et al. 1995) where $b$ = 9.72$^o$,
130~pc $\rm cm^{-3}$ (Taylor \& Cordes 1993), or 229~pc $\rm cm^{-3}$
(Cordes \& Lazio 2002).  These DM models are uncertain, but factors in the range of 1.5 to 3
are larger than the known differences for sources within $\mid b \mid$ $<$ $10\fdeg$ of 
the Galactic plane.  The unusual binary nature of PSR~B1718$-$19 gives support to its association
with NGC~6342 due to its existence as a eclipsing low-mass binary pulsar (Freire 2005) and 
due to the fact that the proportion of pulsars with binary companions is about 
two orders of magnitude greater for GC pulsars than Galactic field pulsars 
(Lyne et al. 1995).  In the following sections, given the lack of clear evidence for or against
the association, the results are discussed both with and without
the inclusion of PSR~B1718$-$19 and will be explicitly stated when PSR~B1718$-$19 is not included.
 
\begin{deluxetable*}{llrrrrrccr}[b]
\tablecaption{The four young pulsars currently known in GCs\label{tab:sample}}
\tablecolumns{10}
\tablehead{
  \colhead{PSR}  &  
  \colhead{Cluster}  & 
  \colhead{$P$}  &  
  \colhead{$\dot{P}$}  &   
  \colhead{DM} &
  \colhead{$B$} &
  \colhead{$\tau_c$} &
  \colhead{$S_{1400}$} &
  \colhead{Binary?} &
  \colhead{Timing Solution} \\
  \colhead{}  &  
  \colhead{}  & 
  \colhead{(ms)}  &  
  \colhead{(s~s$^{-1}$)}  &   
  \colhead{(pc $\rm cm^{-3}$)} &
  \colhead{(G)} &
  \colhead{(yrs)} &
  \colhead{(mJy)} &
  \colhead{} &
  \colhead{Reference} 
  }
\startdata
B1718$-$19 & NGC 6342 & 1004 & $1.6 \times 10^{-15}$ & 75.7 & $1.3 \times 10^{12}$ & $9.8 \times 10^6$ & 0.30 & Yes & Lyne et al. (1993)\\
B1745$-$20 & NGC 6440 & 288 & $4.0 \times 10^{-16}$ & 219.4 & $3.4 \times 10^{11}$ & $1.1 \times 10^7$ & 0.37 & No & Freire et al. (2008)\\
J1823$-$3021B & NGC 6624 & 379 & $3.0 \times 10^{-17}$ & 86.9 & $1.1 \times 10^{11}$ & $2.0 \times 10^8$ & 1.04 & No & Lynch et al. (2011)\\  
J1823$-$3021C & NGC 6624 & 406 & $2.2 \times 10^{-16}$ & 86.7 & $3.0 \times 10^{11}$ & $2.9 \times 10^7$ & 0.71 & No & Lynch et al. (2011)
\enddata
\end{deluxetable*}

\section{Sensitivity limits for young pulsars in GCs}\label{sec:fluxlimits} 

To carry out the statistical analysis of this sample, it is necessary to
have a substantial compilation of upper limits from GCs that have been searched
for such pulsars. Table 2 gives our compilation of
flux density limits at 1400~MHz for searches of each cluster, $S_{\rm min}$ and
GC properties relevant to this work.  
The main surveys that were used are Hessels et al.~(2007), Lynch \& Ransom~(2011), 
Lynch et al.~(2011), and Possenti et al.~(2010). 
The rest of the GCs had their flux limits taken  
from the most recent discovery paper.  If the paper did not quote a flux limit,
one was derived using the survey parameters from the paper using the radiometer 
equation for pulsars which gives

\begin{equation}
S_{\rm min} = \frac{\beta(\rm S/N_{min})T_{\rm sys}}{G\sqrt{n_{\rm p}t_{\rm int}\Delta f}}\sqrt{\frac{\delta}{1-\delta}},
\label{eq:rad}
\end{equation}

\noindent where $\beta$ is a digitization correction faction, $\rm S/N_{min}$ is the 
minimum signal-to-noise ratio, $T_{\rm sys}$ is the
system temperature of the telescope, $G$ is the gain of the telescope, $n_{\rm p}$ is the 
number of polarizations summed, $t_{\rm int}$ is the observation time, $\Delta f$ is
the bandwidth of the backend, and $\delta$ is the fractional pulse width 
(Lorimer \& Kramer 2005).  

The parameters $t_{\rm int}$, $\Delta f$, and $n_{\rm p}$ are all taken from the relevant survey
paper.  The values for $G$ and $\beta$ can be obtained from the websites or papers that give the telescope 
and backend specifications.  $\rm S/N_{min}$ and $\delta$ are parameters that are chosen 
to have values of 8 and 0.1 respectively.  $T_{\rm sys}$ can be expressed as: 

\begin{equation}
T_{\rm sys} = T_{\rm rec} + T_{\rm CMB} + T_{\rm gsyn} + T_{\rm spill} + T_{\rm RFI},
\label{eq:temp}
\end{equation}

\noindent where $T_{\rm rec}$ is the receiver temperature, $T_{\rm CMB}$ is the cosmic background
temperature, $T_{\rm gsyn}$ is the Galactic synchrotron temperature, $T_{\rm spill}$ is the 
spillover temperature from sources in the side lobes of the telescope, and $T_{\rm RFI}$ 
is the increase in system temperature due to terrestrial radio frequency interference (RFI).
The values of $T_{\rm rec}$, $T_{\rm CMB}$, and $T_{\rm gsyn}$ are well known, but $T_{\rm spill}$
and $T_{\rm RFI}$ can vary greatly with time and telescope position causing large uncertainties 
in $T_{\rm sys}$ and can only truly be obtained by proper calibration.  
Due to pulsar survey observations rarely ever having accurate calibration, the expression for 
$T_{\rm sys}$ has been simplified to include only the $T_{\rm rec}$, $T_{\rm CMB}$, 
and $T_{\rm gsyn}$.  One last fact to mention is that, due to these pulsars having long
periods, DM smearing and scattering are unlikely to be important and are not considered in this work.

For any surveys
which were not carried out at 1400~MHz, the quoted limits are scaled from the
observing frequency to 1400~MHz using a simple power law $S \propto \nu^{-1.6}$,
which is consistent with the average spectral behavior for a large sample of normal
pulsars (Lorimer et al. 1995).

\section{The potentially observable population of young pulsars in GCs} \label{sec:statistics} 

For all clusters listed in Table~\ref{table:gclim}, we can use the flux density limits
to model the most likely number of potentially observable 
pulsars\footnote{The estimates in this section do not account for beaming 
effects, or the fact that many pulsars will escape the cluster potential and
will be missed by the surveys. We discuss these issues in \S \ref{sec:intrinsic}.}
in each cluster.
Using a binomial data model, we carry out a simple Bayesian analysis 
described below.
This method rests on a key simplifying assumption about the luminosity
function of young pulsars. Following Faucher-Gigu\`ere \& Kaspi (2006), we
will assume that the parent population follows a log-normal luminosity
function defined (at 1400~MHz) to have a mean in the base-10 logarithm
of $L$ (in mJy) to be $-$1.1 and a standard deviation of 0.9. This assumption is
reasonable if we consider that the spin-down evolution of isolated pulsars in
GCs is the same as the Galactic disk. Variations on the mean and
standard deviation of the log-normal luminosity function have been applied 
to our analysis using models from Ridley \& Lorimer (2010).  These values are in 
the range of $-$1.04 to $-$1.19 for the mean of the luminosity function in the base-10
and in the range of 0.91 to 0.98 for the standard deviation.  

We can model the number of 
pulsars in a particular cluster, $N$, given a ``detection probability''
$\theta$ and an observed sample of $n$ pulsars.
Bayes' theorem gives the joint
posterior probability density for $N$ and $\theta$ as
\begin{equation}
  p(N,\theta|n) \propto p(n|N,\theta) p(N,\theta).
\end{equation}
Here $p(n|N,\theta)$ is the probability of
observing $n$ pulsars from a parent population of $N$ with some $\theta$,
and $p(N,\theta)$ is the joint prior probability density for $N$ and $\theta$. 
The prior distribution for $N$ is assumed to be uniform
in the range $n$ to $\infty$. Graphically, $\theta$ is 
the ratio of the area under the
luminosity function for $L>L_{\rm min}$ to the total area under the
function. We evaluate $\theta$ numerically using Monte Carlo integration of
the Faucher-Gigu{\`e}re \& Kaspi (2006) luminosity function and four
luminosity functions from Ridley \& Lorimer (2010). 
The prior distribution for $\theta$ is assumed to be independent of $N$ and
uniform in the range $\theta_{\rm min}$
to $\theta_{\rm max}$, which are defined to be the minimum and
maximum probabilities from the five distribution functions.  
The simplest choice for a likelihood function is the binomial 
distribution, i.e.
\begin{equation}
  p(n|N,\theta) = \frac{N!}{n!(N-n)!} \theta^n (1-\theta)^{N-n},
\end{equation}
where $\theta$ depends on the
luminosity limit and the assumed parent luminosity
function.  

\begin{figure*}[t]
\epsscale{1.}
\plotone{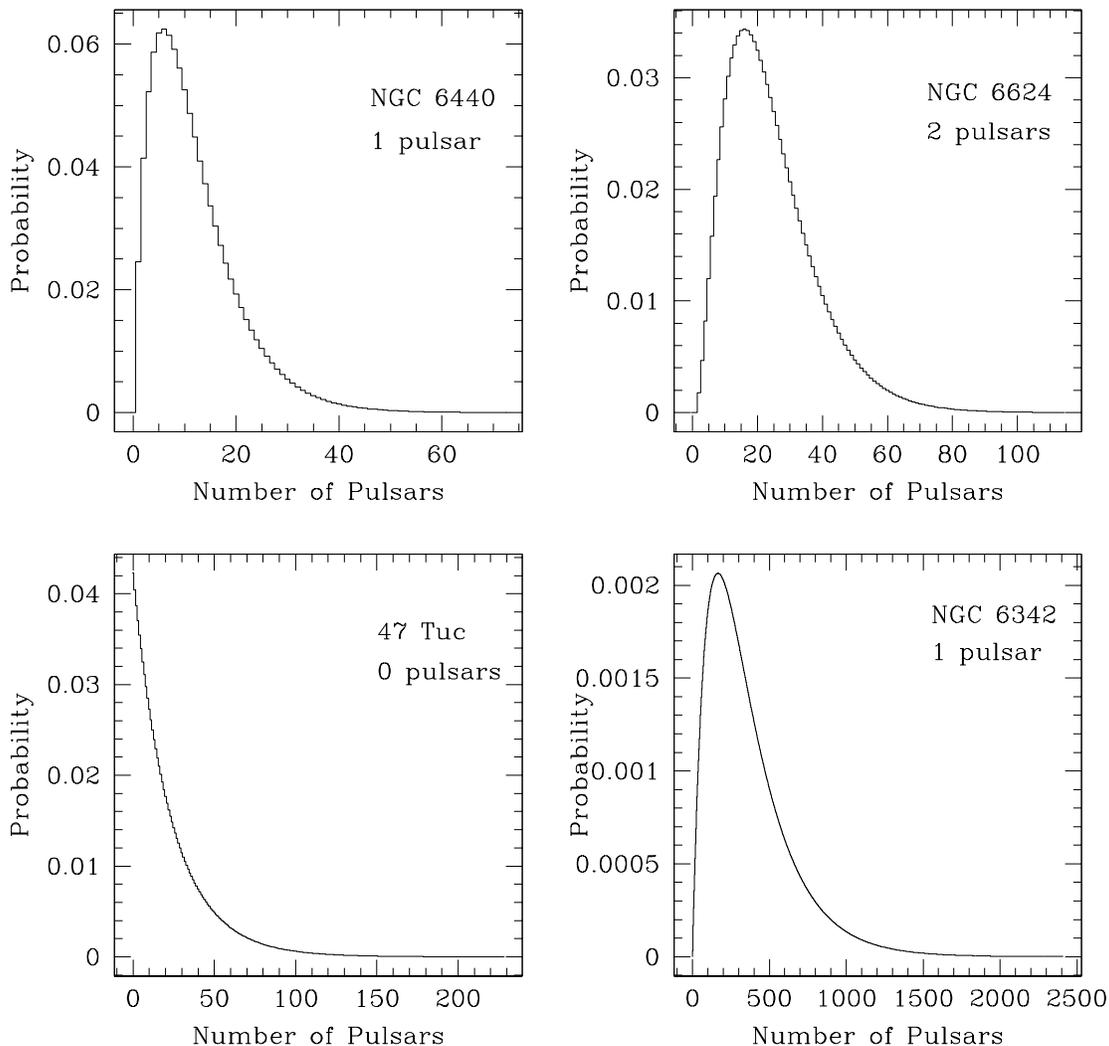}
\caption{\label{fig:dpf}
Examples of discrete posterior probability density functions for the number 
of potentially observable pulsars in GCs using the Bayesian analysis. }
\end{figure*}

For the majority of cases in which there are no detections in a 
cluster, $n=0$ and the likelihood term simplifies considerably to
\begin{equation}
\label{equ:n0}
  p(n|N,\theta) = (1-\theta)^{N}.
\end{equation}
For the cases of NGC 6440 and NGC 6342, where there is one 
detection, i.e. $n=1$, we have
\begin{equation}
\label{equ:n1}
  p(n|N,\theta) = N\theta(1-\theta)^{N-1},
\end{equation}
while for the two pulsars in NGC 6624, $n=2$ which leads to 
\begin{equation}
\label{equ:n2}
  p(n|N,\theta) = \frac{N(N-1)}{2}\theta^2(1-\theta)^{N-2}.
\end{equation}
Having found the joint posterior distribution $p(N,\theta|n)$,
we then marginalize over $\theta$ to get the
posterior distribution for $N$ using the appropriate
choice for the likelihood function (i.e.~Eq.~\ref{equ:n0},
\ref{equ:n1}, or \ref{equ:n2} depending on the value for
$n$ in each case) and give the 95\% percentile-based
credible intervals for $N$ (i.e.~the 0.025 and 0.975 percentiles) as
well as the median in Table~\ref{table:gcbin}.  Examples of these
discrete probability density functions can be seen in 
Figure~\ref{fig:dpf} for four GCs.  47~Tuc shows the typical shape of 
a discrete probability density function for a GC with no young pulsars.

\section{The intrinsic population and birth rates of young GC pulsars} \label{sec:intrinsic} 

The results from the previous sections do not take into account the population
of young pulsars whose emission beams do not intersect our line of sight or the 
population of young pulsars which escaped the gravitational potential of their parent
GCs.  Both of these issues are addressed in this section.  

\subsection{Results for all GCs} 

Retention fractions ($f_R$), the fraction of pulsars which do not have a large 
enough birth velocity to escape the cluster's gravitational potential, are calculated 
for each GC.  These $f_R$ are obtained by using the escape velocity of the GC and a velocity 
distribution function.  Hobbs et al. (2005) showed that a Maxwellian distribution fits
the Galactic population of pulsars well for many categories (all, young, recycled, etc.) 
of pulsars, hence we choose the 
velocity distribution function to have the form of a Maxwellian
with a dispersion $\sigma_{v}$.  Caution does need to be taken with this choice of velocity 
distribution function because no physical mechanism is presented in Hobbs et al. (2005) to explain 
this Maxwellian distribution and the low velocity end of this distribution is not well
constrained due to the fact low velocity pulsar's proper motions are difficult to measure.  
Multiple values were chosen for $\sigma_{v}$:  
265 km~s$^{-1}$ from Hobbs et al. (2005), an intermediate value of 130 km~s$^{-1}$, 50 km~s$^{-1}$,
20 km~s$^{-1}$, and 10 km~s$^{-1}$  as a lower value.  Many separate values have been chosen 
because it is most likely that these young GC pulsars are formed from electron capture 
supernovae for which the accompanying natal kick may be 10 times smaller than in the case of core 
collapse supernova (Ivanova et al. 2008, Kitaura et al. 2006).  The escape velocities were 
taken from Gnedin et al. (2002) except for ESO 452 which came from Webbink (1985).  
The retention fractions are calculated by numerically integrating the Maxwellian 
velocity distribution function from 0 to the escape velocity of the GCs.

Each retention fraction was used to calculate a theoretical upper limit
for the number of young pulsars produced for every GC

\begin{equation}
N_{\rm created} = \frac{N}{f_R f_{\rm beam}},
\label{eq:bir}
\end{equation}

\noindent where $N$ represents the number of pulsars predicted by the 
binomial method, i.e the median in Table~\ref{table:gcbin}, and $f_{\rm beam}$ 
is the beaming fraction for pulsars.   
The value taken for the pulsar beaming fraction is 0.1 (Tauris \& Manchester 1998).  
This value represents an upper limit on the number of pulsars created in a particular GC.  
The number of pulsars created in a GC is then divided by the average life time 
of a young pulsar (43 Myr) to obtain upper limits on the birth rates (${\cal R}$) for each GC.  
An average lifetime of 43 Myr is derived by taking the total number of Radio-Loud pulsars 
(1,200,000) and dividing it by the pulsar birthrate (2.8~psrs per century) in FK06.
All results from the binomial analysis are contained within Table~\ref{table:gcbin}.   

\begin{figure}[t]
\epsscale{1.}
\plotone{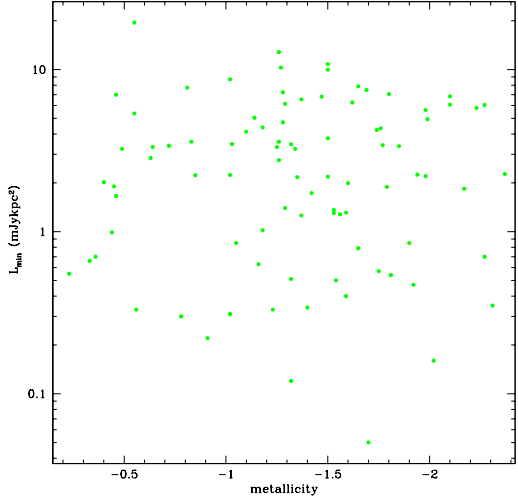}
\caption{\label{fig:lim}
1400 MHz luminosity survey limits as a function of metallicity with six GCs excluded with 
distances greater then 20 kpc.  The luminosities are randomly distributed and show 
no bias against low or high metallicity GCs.}
\end{figure}

One GC, M22, has median value of zero and no young pulsars are expected to be 
contained within it.  This results in a birth rate of zero for M22.  To calculate
a birth rate for this GC, it is assumed that the cluster contains one pulsar
and an upper limit is constructed using this assumption.

\subsection{Metal-rich GCs} 

Up to this point, the analysis presented uses flux density limits and luminosity
models as the only considerations for observable pulsars contained within the cluster.  
Lyne et al.~(1996) note that the young pulsars appear in metal-rich
GCs and this trend has persisted despite fifteen years of intense searches
of most of the cluster population.  In this section metallicity 
will also be added as a consideration.  For the following discussion Terzan~5 will 
be excluded due to the possibility that it is not a true GC but a merger of two 
astrophysical objects that are bound in the Galactic halo (Ferraro et al. 2009).  
Also excluded will be B1718$-19$ in NGC~6342 due to the uncertainty about its 
membership to its cluster (see \S \ref{sec:sample}).  

The GCs NGC~6440 and NGC~6624 have metallicities that are greater than the $90^{\rm th}$ 
percentile of metal rich clusters.  These two clusters are two of the three highest metallicity 
clusters that contain either young or old millisecond pulsars.  The probability of selecting 
two GCs in the top three of a ranked list from a sample of 25 GCs is 0.92 \%.  
If NGC~6342 is also included in this
sample then we have three of the top five highest metallicity clusters with a  
probability of selecting three of the top five of 0.38 \%.  The inclusion of Terzan~5 (the highest 
metallicity cluster with any known pulsars) in both 
of these previous scenarios, changes the probability values from 0.92 \% to 1.71 \% 
and 0.38 \% to 0.68 \%.  The conjecture can be proposed that 
metal-poor clusters have been selected against for the purpose of surveying GCs due to 
the belief that metal-rich GCs contain more pulsars overall.  Figure~\ref{fig:lim} shows 
a plot of metallicity versus survey luminosity limit, indicating a random 
distribution with no bias towards either low or high metallicity clusters.  

\begin{figure}[t]
\epsscale{1.}
\plotone{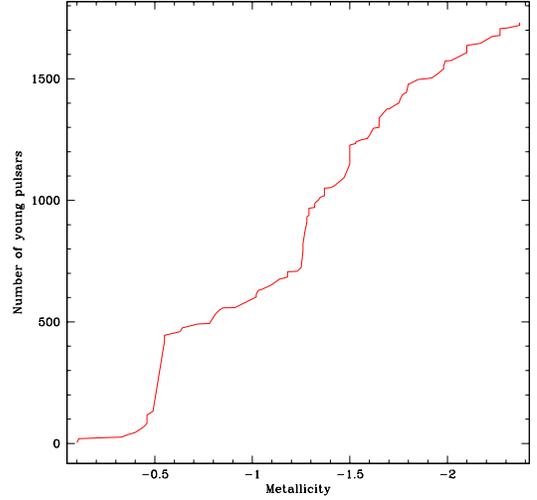}
\caption{\label{cdf:met}
CDF of young pulsars predicted by the binomial
analysis as a function of metallicity.}
\end{figure}

Figure~\ref{cdf:met} shows the empirical cumulative distribution function (CDF) of the 
number of predicted young pulsars with the binomial method versus metallicity.  GCs 
with distances greater than 20 kpc are excluded from this CDF because their large 
distances produce high luminosity limits and in turn creates jumps in the CDF.  
The cause of these jumps seen in Figure~\ref{cdf:met} is due to high survey limits 
on specific GCs. The first large jump in the CDF at a metallicity of $\sim-$0.55
is due to NGC~6342, the GC with the highest survey luminosity limit.

The GCs with the most known pulsars, Ter~5 and 47~Tuc, have high metallicities and 
have been used in many simulations.  One of these presented in Ivanova et al. (2008) 
attempts to model the number of young/high magnetic field pulsars in these clusters.  
In these simulations, 3 and 2 young pulsars are predicted to exist in the cores of Ter~5 
and 47~Tuc respectively with the most likely creation scenario being the merger of two stars.  
If beaming and luminosity limits from Table~\ref{table:gclim} are taken into account, 
the chances of seeing those 2-3 pulsars are extremely small.

\section{Discussion}\label{sec:discussion} 

\subsection{Flux Luminosity Limits} 

The results of the binomial analysis depend greatly on the luminosity limits provided 
from searching these GCs.  One can see by comparing the discrete probability density functions of 
NGC~6342 and NGC~6440 seen in Figure~\ref{fig:dpf}, that the range of values that each function
covers differs greatly.  This is a direct result of NGC~6342's flux luminosity limit being 
much greater than that of NGC~6440.  Thus clusters with high flux luminosity limits do not
significantly constrain the population/birth rates of these GCs.  

Another factor not mentioned is the influence of RFI on searching GCs. The relative 
impact of RFI on long-period pulsars is much more severe than for millisecond pulsars 
(MSPs).  There is a lot more long-period RFI and the dispersion discrimination between RFI 
and long-period pulsars is not nearly as great as for MSPs.  Therefore this creates 
a bias against finding such pulsars because they may be ignored assuming they are RFI 
which is hard to quantify and fully account for.

\begin{figure}[t]
\epsscale{1.}
\plotone{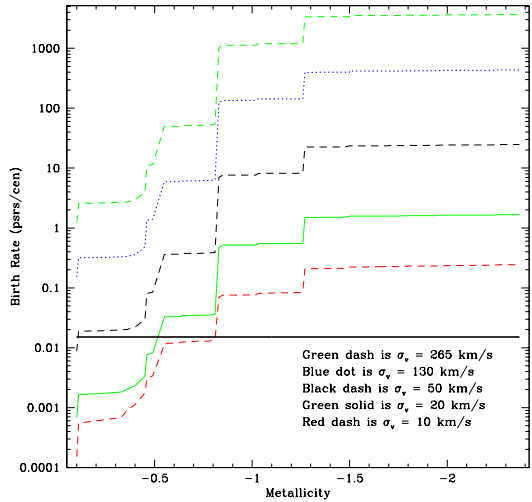}
\caption{\label{cdf:bir}
CDF of birth rate versus metallicity.}
\end{figure} 

\subsection{Birth Rates in GCs} 

For most GCs, the birth rates found are only upper limits with the exception of 
NGC~6342, NGC~6440, and NGC~6624.  Empirical CDFs of these values can be seen in 
Figure~\ref{cdf:bir} for the 97 GCs surveyed.  The upper limits on birth rates for 
velocity dispersions of 265, 130, 50, 20, and 10 km~s$^{-1}$ are 3568, 422, 24.8, 1.67, 
and 0.25 pulsars per century respectively.  The higher birth rates obtained from the 
higher velocity 
dispersions is an effect of needing to produce more pulsars to get enough pulsars at the
low velocity end of the Maxwellian distribution.   These values for birth rates provided 
in this work are much higher then the predicted birth of 2.8 pulsars per century for 
the Galaxy as a whole for larger velocity dispersions (Faucher-Gigu{\`e}re \& Kaspi 2006).  
The impact on the Galactic 
population from pulsars escaping GCs will be discussed elsewhere (Lynch et al., in prep). 
For the purposes of this paper, these very high implied birth rates suggest that a very 
different formation process for young pulsars is occurring in GCs, as well as the 
possibility that some GCs do not produce young pulsars at all.  The major difference in the 
formation scenario appears to be the lower velocity dispersion at birth.  

\subsection{Role of cluster metallicity} 

\begin{figure}[t]
\epsscale{1.}
\plotone{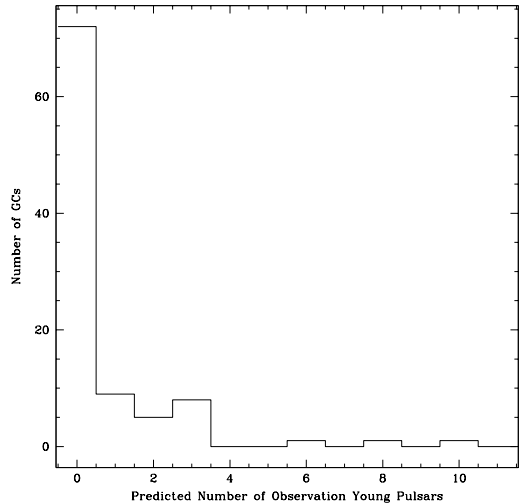}
\caption{\label{his:mass}
Histogram of number of predicted observable young pulsars per Globular Cluster using 
a model only dependent on the mass of the GC.}
\end{figure}

In Section 5.2, evidence is presented that young pulsars are only present in higher
metallicity GCs.  
Another way to highlight this is to use a simple population model with the GC's mass as 
the only variable.  A pulsar to mass ratio (PMR) is derived using the results of the 
binomial analysis for NGC 6440 and NGC 6624, each having 10 and 21 pulsars predicted 
respectively.  We do not include the GC NGC 6342 when creating this model
because of the uncertain cluster membership of B1718$-$19 
(see \S \ref{sec:sample}).  With NGC 6440's mass of 811,000 $\msun$ and NGC 6624's mass of 
257,000 $\msun$ a PMR of 31 pulsars per 1,068,000 $\msun$ is obtained.   Using the PMR each 
of the 97 GCs are revisited and an intrinsic population is predicted for each GC by multiplying the PMR by the GC mass.
Using the flux density limits in Table~\ref{table:gclim}, the observable population 
is drawn from the Faucher-Gigu\`ere \& Kaspi (2006) luminosity
model.   The results of this simulation can be seen in a histogram in
Figure~\ref{his:mass} and Table~\ref{table:gcbin}.  Seventy-four percent of the 
GCs are predicted to have no observable
young pulsars using this model.  However, for the remaining GCs, a total of 67 young 
pulsars should be observable which
disagrees with the current population of 3 by over an order of magnitude.  These results
show that mass is not a single determining factor in a GC containing young pulsar.

Based on the observed population of young pulsars in the higher metallicity GCs, it may be 
suggested that lower metallicity GCs may not produce any young pulsars. Figure~\ref{cdf:bir} 
may be used to predict birth rates as a function of metallicity. Given that no young 
pulsars are observed with metallicities below $-0.6$, this value is adopted as a cut-off 
value. In this case, the population of young pulsars in GCs with log[Fe/H] $>$ $-0.6$ is 
$447^{+1420}_{-399}$, at $95\%$ confidence level.
This implies an upper limit on the birth rate for GCs with log[Fe/H] $>$ $-0.6$ of
$0.012^{+0.037}_{-0.010}$ pulsars per century.  

\subsection{Formation scenarios}  

A few possible situations need to be examined that could explain this current population
of normal pulsars in globular clusters.  

\subsubsection{Blue Stragglers} 

Blue stragglers are the product of the merger of two or possibly three $\sim$1$\msun$ 
main sequence stars (Leonard, 1989) or binary accretion of a binary companion(s) (McCrea, 1964).  
They cannot be the progenitors of the young pulsars because they are
not massive enough to core collapse.

\subsubsection{Electron Capture Supernovae} 

The creation of these normal pulsars from electron capture supernova (ECS) 
of Oxygen-Neon-Magnesium (ONeMg) white dwarfs is another possible avenue.  
There are three main types of ECSs: 
accretion-induced collapse (AIC),  evolution-induced collapse (EIC), and
merger-induced collapse (MIC), of which AIC is the most common 
(Ivanova et al. 2008).  From conservation of magnetic flux,
a five order of magnitude change in surface magnetic field is available
from an AIC of a white dwarf into a neutron star.  This would require a
white dwarf to have at least $10^7$ Gauss magnetic field to produce
the magnetic field of the young pulsar with the largest magnetic field 
contained within a GC.
White dwarfs have been observed with magnetic field above $2 \times 10^6$
Gauss and could account up to $10\%$ of the white dwarf population
(Liebert et al 2003).  ECSs could be an avenue for the creation of young
pulsars.  Not all white dwarfs that pass the Chandrasekhar limit will 
collapse via ECS, some will be type Ia supernova.  

ECS provide two qualities that are needed to explain the presence of young 
pulsars in GCs.  The first is the low velocity dispersion which is needed
to keep these young pulsars in their host GCs.  The initial formation of the 
white dwarf results in a small velocity kick of a few km/s from either 
stellar winds or an asymmetric kick during the helium flash (Fregeau et al. 2009).  
The energetics of the ECS of the white dwarf would produce a small velocity kick 
(Dessart et al. 2006) and, combined with the velocity kick from previous stages of evolution, 
would provide a total velocity that is small and would allow most neutron stars created 
via ECS to be retained by the GC.  The second is the higher metallicity of GCs which 
host young pulsars.  An ECS ejects a few 0.001 $\msun$ worth of mass, of which $\sim 25 \%$
is $^{56}$Ni (Dessart et al. 2006).   $^{56}$Ni decays into $^{56}$Co with a half-life
of $\simeq 6$ days and then $^{56}$Co decays into $^{56}$Fe with a half-life of $\simeq 77$ days.
This provides a source of iron creating the higher metallicity for clusters in which
all known young pulsars have been detected.

A simple statistical analysis can give an order of magnitude estimate of the 
possible population of heavy mass (1.0--1.4~$\msun$) white dwarfs available 
for ECSs.  It is taken that all GCs have the same initial mass function (IMF)
and the differences seen between current GC's mass functions are due to 
the dynamical history of the cluster (Paust et al. 2010).  The IMF is a multi-part
power-law presented in equation~4 of Kroupa (2002) and with $\alpha_3$ = $-$2.7 
from Scola (1986).  Each GC is assumed to have the same age of 12.4 Gyr for 
purposes of predicting an initial mass of the GC (Krauss \& Chaboyer 2001).  
An initial mass is obtained for each GC by using the mass lost fractions
presented in Kroupa (2002) for $\alpha_3$ = $-$2.7.  The number of main sequence
stars available for heavy white dwarfs (mass range of 6.0 - 8.0 $\msun$) are calculated 
for each GC with a total of $\sim$150,000 white dwarfs formed in these 97 GCs.  The use of 
values for $\alpha_3$ greater then $-$2.7 serves only to decrease the numbers of stars 
in this mass range while choosing the Salpeter index of $\alpha_3$ = $-$2.35 only 
increases the number of stars available by a factor of 2.  Assuming an interaction age 
of $10^9$ years and that each heavy white dwarf creates one ECS, an occurrence rate of 
$1.5 \times 10^{-2}$ ECSs per century is obtained and is shown on Figure~\ref{cdf:bir} 
as a solid horizontal black line. 

This predicted ECS rate is an order of magnitude less than the birth rates of pulsars
for the lowest velocity dispersion.   A few factors could bring these values into closer 
agreement.  A decrease in the interaction age would increase the ECS rate.  This is a very 
plausible possibility because there is an accumulation of ECSs at more recent times due to 
dynamical evolution (it may take a long time for a given capture to form an accreting binary)
and individual evolution (it may take several Gyrs to accrete enough mass to cause an ECS).
Lower luminosity
limits without the detection of more young pulsars would decrease the young pulsar birth rate
and bring the two rates closer to agreement.  A detailed $N$-body simulation should be used to refine
the ECS rate but this is beyond the scope of this work.  It is possible that ECSs have only 
occurred in higher metallicity cluster and not all GCs should be considered in calculating the 
birthrate of young pulsars.  The work presented here suggests that ECS is the most likely 
creation scenario for young pulsars in GCs.   

\begin{deluxetable*}{lrrcccrrc}[b]
\tablecaption{Parameters for Bulge Globular Clusters. \label{tab:gcbul}}
\tablecolumns{9}
\tablehead{
  \colhead{Globular}  &
  \colhead{$l$}  &
  \colhead{$b$}  &
  \colhead{Distance}  &
  \colhead{Dist. from Gal. center} &
  \colhead{$r_t$} &
  \colhead{$S_{\rm min}$} &
  \colhead{Mass} &
  \colhead{$V_{\sigma}$} \\
  \colhead{Cluster}  &
  \colhead{($^o$)}  &
  \colhead{($^o$)}  &
  \colhead{(kpc)}  &
  \colhead{(kpc)} &
  \colhead{(pc)} &
  \colhead{(mJy pc$^2$)} &
  \colhead{($\msun$)} &
  \colhead{(km~s$^{-1}$)}
  }
\startdata
NGC 6325 &   0.97 &  8.00   & 7.8 & 1.1 & 15.8 & 54.8 & 223,000 & 5.9  \\
NGC 6355 & 359.59 &  5.43   & 9.2 & 1.4 & 31.6 & 77.5 & 252,000 & N/A  \\
Terzan 2 & 356.32 &  2.30   & 7.5 & 0.8 & 24.0 & none & 3,290 & N/A  \\
Terzan 4 & 356.02 &  1.31   & 7.2 & 1.0 & 23.3 & none & N/A & N/A  \\
HP 1     & 357.44 &  2.12   & 8.2 & 0.5 & 20.4 & none & 95,700 & N/A  \\
Liller 1 & 354.84 & $-$0.16 & 8.2 & 0.8 & 36.5 &  9.9 & 289,000 & N/A  \\
Terzan 1 & 357.57 &  1.00   & 6.7 & 1.3 & 22.8 & 77.5 & 5,360 & N/A  \\
Ton 2    & 350.80 & $-$3.42 & 8.2 & 1.4 & 25.3 & none & 7,330 & N/A \\
Terzan 5 &   3.84 &  1.69   & 6.9 & 1.2 & 21.9 & 11.6 & 374,000 & N/A  \\
NGC 6440 &   7.73 &  3.80   & 8.5 & 1.3 & 13.0 &  9.8 & 811,000 & N/A  \\
Terzan 6 & 358.57 & $-$2.16 & 6.8 & 1.3 & 43.6 &  7.3 & 300,000 & N/A  \\
UKS 1    &   5.13 &  0.76   & 7.8 & 0.7 & 45.6 & 54.8 & 145,000 & N/A  \\
Terzan 9 &   3.61 & $-$1.99 & 7.1 & 1.1 & 21.3 & 17.0 & 9,570 & N/A  \\
NGC 6522 &   1.02 & $-$3.93 & 7.7 & 0.6 & 34.8 & 54.8 & 300,000 & 6.7  \\
NGC 6528 &   1.14 & $-$4.17 & 7.9 & 0.6 & 33.1 & 54.8 & 152,000 & N/A  \\
NGC 6558 &   0.20 & $-$6.02 & 7.4 & 1.0 & 28.5 & 63.3 & 98,400 & 3.1  \\
NGC 6624 &   2.79 & $-$7.91 & 7.9 & 1.2 & 47.4 & 16.0 & 257,000 & 5.4  \\
\enddata
\end{deluxetable*}

\subsubsection{Galactic Bulge Pulsars} 

One other possibility is that these young pulsars are part of the Galactic field population
and are captured by their host GC.  The time it would take to travel the distances to the GCs 
and the time it would take for these pulsars to relax into the core of their host GC have been 
proposed as evidence against this method.  One key fact is neglected here; these are all bulge 
GCs that contain this population of young pulsars.  Their distances from the Galactic center 
range from 1.2--1.7~kpc and the radius of the Galactic bulge is $\sim 1.5$~kpc.  The time for 
a pulsar to travel from the outer 0.5 kpc of the Galactic bulge would be short ($\sim 10^6$ years) 
even for a moderate birth velocity kick and the core relaxation time for each GC is an order 
of magnitude less than the characteristic age of the pulsars that hosts them allowing enough
time for them to settle into the core which is where 3 of the 4 pulsars are found.  The fourth, 
PSR B1718$-$19, has a characteristic age about equal to NGC 6342's relaxation time and this could 
explain why PSR B1718$-$19 is not located in the host GC's core.  This could also explain why 
these young pulsars are towards the higher end of the age distribution for non-recycled pulsars.  
The metallicity of the Galactic center is higher than that of the Galactic disk and stars
migrating from the bulge to bulge GCs could explain why GCs with young pulsars have higher 
metallicities compared to the overall GC population.  

The population of bulge pulsars is relatively unexplored.  Outside of Galactic center surveys 
(Deneva et al, 2009; Johnston et al, 2006; Klein et al, 2004; \& Kramer et al, 2000) and 
the surveys of bulge GCs (Table~\ref{table:gclim}), the population is unknown.  The most
sensitive published survey to explore this region is the Parkes Multi-beam survey 
(Manchester et al, 2001).  For pulsars with a period of 100-1000 ms the survey has a 
limiting flux of 0.16 mJy and this translates into a $L_{\rm min}$ of 7.8 mJy$\rm kpc^2$
for a distance of 7.0 kpc (front edge of bulge) and 12 mJy$\rm kpc^2$ for a distance of
8.6 kpc (top and bottom of bulge directly above and below Galactic center).  These 
$L_{\rm min}$ values represent the upper edge of the pulsar luminosity function presented
in FK06 and show that the Galactic bulge has not been surveyed well enough to constrain
its pulsar population especially considering DM smearing and scattering would
further hinder the detection of these pulsars.  

Another chance for pulsars to be captured 
will occur when the GC passes through the plane of the Galaxy as it orbits the Galactic center.
The orbital timescales for globular clusters around the galaxy are hundreds of Myrs and
this is longer than the 43 Myr average lifetime of a normal pulsar. 
This means that any pulsars picked up would have to have been picked up on the most recent pass
of the GC through the Galactic disk.  Figure~\ref{fig:ppdot} shows a small population of pulsars
that would live long enough to be found in a GC if capture occurred due to this scenario.  

The plausibility of star capture by a GC is discussed in Mieske \& Baumgardt  
(2007; henceforth MB07) and their arguments will be applied here in the following discussions.  
Table~\ref{tab:gcbul} lists all GCs within 1.5~kpc of the Galactic center and their 
characteristics relevant to the MB07 analysis.  MB07 found 
that the capture probability decreases with increasing number of cluster particles ($N_{\rm GC}$)
and decreases with increasing initial velocity over cluster velocity dispersion ratio 
($\frac{V_{\rm int}}{\sigma_{\rm GC}}$).  Assuming an average particle mass of 0.5~$\msun$, the range
of clusters reviewed matches with the range of Galactic bulge clusters, however the range
of $\frac{V_{\rm int}}{\sigma_{\rm GC}}$ do not.  GC velocity dispersions range from $\sim$ 
1.0 - 19~km~s$^{-1}$, which are an order of magnitude less than typical pulsar velocities in previous
studies (see Hobbs et al. 2005, for a recent study).  The closest scenario of 
$\frac{V_{\rm int}}{\sigma_{\rm GC}}$ equal to unity will be the only one considered from here.

To obtain a comparison of capture rates between pulsars and MB07, a few other considerations
need to be put into place.  Using the FK06 model for the Galactic pulsar population, the mass
density of pulsars is found to be 8.2$\times 10^{-6}$ $\msun$ $\rm pc^{-3}$ for the Galactic 
bulge assuming a uniform pulsar per mass distribution throughout the 
Galaxy.  This value is five orders of magnitude less than the value of 0.25 $\msun$ $\rm pc^{-3}$
used in MB07.  Even if the entire pulsar population were placed inside the Galactic bulge, this would
increase the mass density by only an order of magnitude.  The resulting differences in mass density
would produce significantly lower values for the rates found by MB07.  The value of 
265~km~s$^{-1}$ will be adopted as the velocity dispersion for field pulsars 
($\sigma_{\rm field}$).  For a $\sigma_{\rm field}$ of
200~km~s$^{-1}$ MB07 concludes that no stars will be captured in any mass cluster within a 
Hubble time.  If the additional constraints for pulsars of lower capture probability due to
higher initial velocities and lower capture rates due to lower mass density are included, the
conclusion can be drawn that no pulsars are likely to be captured by a GC.  Imposing further
constraints of beaming, luminosity limits, and finite radio-loud lifetimes for pulsars would 
further hinder the detection of a pulsar if one were to be captured by a GC.   We therefore
rule out this possibility as an origin for the young pulsars in GCs.

\subsection{Suggested Future Work}  

All the work presented here only provides a statistical study at the young pulsar 
population in GCs and neglects the dynamics and history of each GC.  Clearly N-body
simulations similar to those presented in Ivanova et al. (2008) of the GCs and their 
possible interactions with the Galactic stellar and post-stellar populations would 
place better constraints on the values presented here and would further the understanding of 
GC's dynamics and evolutionary history.

To improve upon the observational constraints used in this work, it is clearly desirable 
to search all GCs as deeply as possible using existing facilities. We highlight here 
some GCs of particular interest.  

There are a few exceptions, most notably NGC~6342 which has the highest minimum
detectable flux density of any of the 97 GCs and has not to our knowledge 
been surveyed since the discovery of B1718$-$19 by Lyne et al.~(1993).
A new search with the currently available
telescopes could reduce the $\rm S_{min}$ by a factor of 10 and provide evidence for
or against PSR~B1718$-$19 association with NGC 6342 (Bailes et al. 2005, Freire  
2005).

Another high metallicity bulge GC to be searched is NGC~6637 (M~69). This cluster, at a distance 
of 8.8 kpc, also has a high two-body encounter rate, $\Gamma$. This parameter is often 
used to assess the plausibility of pulsar content in GCs (see, e.g., Hui et al. 2010).

Most of the GCs presented in this work have been searched to the sensitivity limit of the 
telescope for which they are visible.  The next generation of radio telescopes will be needed
to present better constraints on this work.  The Square Kilometer Array (SKA) would be an ideal
telescope to accomplish this task.  The SKA would provide over a factor of one hundred 
increase in telescope gain over the GBT, Parkes, and Jodrell Bank telescopes and over a factor
of thirty in telescope gain over Arecibo.  This improvement in gain alone would allow for 
deep searches, with luminosity limits less than the mean in the Faucher-Gigu\`ere \& Kaspi (2006) 
model, in only an hour or 
two for GCs less than twenty kpc away (see Smits et al, 2009 for pulsar work with the SKA).  
Before the advent of the SKA, MeerKAT, the South African SKA precursor, will be able to
reduce these minimum detectable flux density limit by a factor of 10 when completed 
(Booth et al, 2009).




\section*{Acknowledgments}\label{sec:awk}
This work was supported by NSF grant AST-0907967 and a WVEPSCoR Research Challenge Grant. 
We would like to thank the scientific editor and anonymous referee for their help 
and comments during the submission process.

\begin{table*}[h]
\begin{center}
\caption{Parameters for globular clusters searched for pulsars.}
\smallskip
\footnotesize
\begin{tabular}{cccccccc}
\hline
Globular & Dist. & $S_{min}$(1400) &  $V_{\rm esc}$ & Metallicity & Mass & Number of & Search \\
Cluster & (kpc) & ($\mu$Jy) & (km~s$^{-1}$) & log[Fe/H] & $\msun$ & Young Pulsars & Reference \\
\hline
47~Tuc & 4.5 & 167.6 & 68.8 & $-$0.72 & 1,500,000 & 0 & Freire et al 2001  \\ 
NGC~288 & 8.9 & 6.5 & 13.3 & $-$1.32 & 112,000 & 0 & Lynch \& Ransom 2011  \\ 
NGC~1261 & 16.3 & 38.8 & 3.4 & $-$1.27 & 341,000 & 0 & Possenti unpublished 2010  \\ 
Pal~2 & 27.2 & 34 & 27.8 & $-$1.42 & 410,000 & 0 & Hessels et al 2007  \\ 
NGC~1851 & 12.1 & 30.1 & 51.8 & $-$1.18 & 551,000 & 0 & Freire et al 2004  \\ 
NGC~2298 & 10.8 & 4.1 & 18.4 & $-$1.92 & 84,900 & 0 & Lynch \& Ransom 2011  \\ 
NGC~2808 & 9.6 & 54.8 & 72.8 & $-$1.14 & 1,420,000& 0 & Possenti unpublished 2010  \\ 
E3 & 8.1 & 54.8 & 3.0 & $-$0.83 & 3,290 & 0 & Possenti unpublished 2010  \\ 
NGC~3201 & 4.9 & 54.8 & 22.0 & $-$1.59 & 254,000 & 0 & Possenti unpublished 2010  \\ 
NGC~4147 & 19.3 & 19 & 18.3 & $-$1.80 & 74,700 & 0 & Hessels et al 2007  \\ 
NGC~4372 & 5.8 & 54.8 & 21.5 & $-$2.17 & 329,000 & 0 & Possenti unpublished 2010  \\ 
NGC~4590 & 10.3 & 54.8 & 18.2 & $-$2.23 & 223,000 & 0 & Possenti unpublished 2010  \\ 
NGC~4833 & 6.6 & 77.5 & 31.8 & $-$1.85 & 410,000 & 0 & Possenti unpublished 2010  \\ 
NGC~5024 & 17.9 & 19 & 33.4 & $-$2.10 & 826,000 & 0 & Hessels et al 2007  \\ 
NGC~5053 & 17.4 & 20 & 8.9 & $-$2.27 & 125,000 & 0 & Hessels et al 2007  \\ 
NGC~5139 & 5.2 & 48.3 & 60.4 & $-$1.53 & 3,350,000 & 0 & Possenti unpublished 2010  \\ 
NGC~5272 & 10.2 & 21 & 37.2 & $-$1.50 & 957,000 & 0 & Hessels et al 2007  \\ 
NGC~5286 & 11.7 & 54.8 & 52.6 & $-$1.69 & 713,000 & 0 & Possenti unpublished 2010  \\ 
NGC~5466 & 16.0 & 22 & 9.5 & $-$1.98 & 179,000 & 0 & Hessels et al 2007  \\ 
Pal~5 & 23.2 & 32 & 3.2 & $-$1.41 & 30,000 & 0 & Hessels et al 2007  \\ 
NGC~5897 & 12.5 & 5.5 & 13.4 & $-$1.90 & 211,000 & 0 & Lynch \& Ransom 2011  \\ 
NGC~5904 & 7.5 & 25 & 47.7 & $-$1.29 & 857,000 & 0 & Hessels et al 2007  \\ 
NGC~5927 & 7.7 & 54.8 & 33.9 & $-$0.49 & 338,000 & 0 & Possenti unpublished 2010  \\ 
NGC~5946 & 10.6 & 54.8 & 25.3 & $-$1.29 & 281,000 & 0 & Possenti unpublished 2010  \\ 
NGC~5986 & 10.4 & 3.7 & 37.0 & $-$1.59 & 599,000 & 0 & Lynch \& Ransom 2011  \\ 
M~80 & 10.0 & 5.7 & 48.7 & $-$1.75 & 502,000 & 0 & Lynch et al 2011  \\ 
NGC~6121 & 2.2 & 131.4 & 34.2 & $-$1.16 & 195,000 & 0 &  Lyne et al 1988  \\ 
ESO452 & 8.3 & 54.8 & 5.9 & $-$1.50 & 75,000 & 0 & Possenti unpublished 2010  \\  
NGC~6144 & 8.9 & 54.8 & 14.1 & $-$1.76 & 169,000 & 0 & Possenti unpublished 2010  \\ 
NGC~6139 & 10.1 & 77.5 & 59 & $-$1.65 & 566,000 & 0 & Possenti unpublished 2010  \\ 
NGC~6171 & 6.4 & 54.8 & 25 & $-$1.02 & 182,000 & 0 & Possenti unpublished 2010  \\ 
NGC~6205 & 7.1 & 27 & 39.1 & $-$1.53 & 775,000 & 0 & Hessels et al 2007  \\ 
NGC~6218 & 4.8 & 54.8 & 28.5 & $-$1.37 & 217,000 & 0 & Possenti unpublished 2010  \\ 
NGC~6235 & 11.5 & 54.8 & 16.8 & $-$1.28 & 73,300 & 0 & Possenti unpublished 2010  \\ 
NGC~6254 & 4.4 & 66.6 & 29.5 & $-$1.56 & 252,000 & 0 & Possenti unpublished 2010  \\ 
Pal~15 & 45.1 & 38 & 4.3 & $-$2.07 & 40,300 & 0 & Hessels et al 2007  \\ 
NGC~6266 & 6.8 & 22.1 & 97.8 & $-$1.18 & 1,220,000 & 0 & Chandler 2003  \\ 
NGC~6273 & 8.8 & 54.8 & 58.4 & $-$1.74 & 1,100,000 & 0 & Possenti unpublished 2010  \\ 
NGC~6284 & 15.3 & 54.8 & 28.6 & $-$1.26 & 361,000 & 0 & Possenti unpublished 2010  \\ 
NGC~6287 & 9.4 & 77.5 & 30.4 & $-$2.10 & 188,000 & 0 & Possenti unpublished 2010  \\ 
NGC~6293 & 9.5 & 54.8 & 41.7 & $-$1.99 & 329,000 & 0 & Possenti unpublished 2010  \\ 
NGC~6304 & 5.9 & 54.8 & 38.4 & $-$0.45 & 217,000 & 0 & Possenti unpublished 2010  \\ 
M~92 & 8.3 & 5.1 & 47.1 & $-$2.31 & 489,000 & 0 & Lynch \& Ransom 2011  \\ 
NGC~6325 & 7.8 & 54.8 & 42.5 & $-$1.25 & 223,000 & 0 & Possenti unpublished 2010  \\ 
NGC~6333 & 7.9 & 54.8 & 37.7 & $-$1.77 & 422,000 & 0 & Possenti unpublished 2010  \\ 
NGC~6342 & 8.5 & 270 & 22.9 & $-$0.55 & 96,600 & 1 & Biggs $\&$ Lyne 1996 \\ 
NGC~6355 & 9.2 & 77.5 & 40.3 & $-$1.37 & 252,000 & 0 & Possenti unpublished 2010  \\ 
Liller~1 & 8.2 & 9.9 & 41.2 & $-$0.33 & 289,000 & 0 & Lynch et al 2011  \\  
Ter~1 & 6.7 & 77.5 & 7.4 & $-$1.03 & 5,360 & 0 & Possenti unpublished 2010  \\ 
NGC~6388 & 9.9 & 54.8 & 124 & $-$0.55 & 2,170,000 & 0 & Possenti unpublished 2010  \\ 
NGC~6402 & 9.3 & 54.8 & 39.1 & $-$1.28 & 1,040,000 & 0 & Possenti unpublished 2010  \\ 
NGC~6401 & 10.6 & 77.5 & 38.3 & $-$1.02 & 286,000 & 0 & Possenti unpublished 2010  \\ 
NGC~6397 & 2.3 & 31.5 & 48.3 & $-$2.02 & 115,000 & 0 & Possenti unpublished 2010  \\ 
Pal~6 & 5.8 & 6.7 & 28.0 & $-$0.91 & 228,000 & 0 & Lynch et al 2011  \\ 
NGC~6426 & 20.6 & 25 & 14.7 & $-$2.15 & 117,000 & 0 & Hessels et al 2007  \\ 
Ter~5 & 6.9 & 11.6 & 50.5 & $-$0.23 & 374,000 & 0 & Ransom et al 2005  \\ 
NGC~6440 & 8.5 & 9.8 & 85.2 & $-$0.36 & 811,000 & 1 & Freire et al 2008  \\ 
NGC~6441 & 11.6 & 12.4 & 102 & $-$0.46 & 1,570,000 & 0 & Freire et al 2008  \\ 
Ter~6 & 6.8 & 7.3 & 38.3 & $-$0.56 & 300,000 & 0 & Lynch et al 2011  \\  
NGC~6453 & 11.6 & 80.4 & 22.4 & $-$1.50 & 169,000 & 0 & Possenti unpublished 2010  \\ 
UKS~1 & 7.8 & 54.8 & 25.4 & $-$0.64 & 145,000 & 0 & Possenti unpublished 2010  \\ 
NGC~6496 & 11.3 & 54.8 & 19.7 & $-$0.46 & 200,000 & 0 & Possenti unpublished 2010  \\ 
Ter~9 & 7.1 & 17.0 & 9.8 & $-$1.05 & 9,570 & 0 & Lynch \& Ransom 2011  \\ 
NGC~6517 & 10.6 & 3.0 & 82.9 & $-$1.23 & 526,000 & 0 & Lynch et al 2011  \\ 
NGC~6522 & 7.7 & 54.8 & 42.3 & $-$1.34 & 300,000 & 0 & Possenti unpublished 2010  \\ 
NGC~6528 & 7.9 & 54.8 & 26.4 & $-$0.11 & 152,000 & 0 & Possenti unpublished 2010  \\ 
NGC~6535 & 6.8 & 41 & 10.0 & $-$1.79 & 20,000 & 0 & Hessels et al 2007  \\ 
NGC~6539 & 7.8 & 47.0 & 35.8 & $-$0.63 & 536,000 & 0 & D' Amico et al 1993  \\ 
\end{tabular}
\smallskip
\label{table:gclim}
\end{center}
\end{table*}

\addtocounter{table}{-1}
\begin{table*}[h]
\begin{center}
\caption{Continued}
\smallskip
\footnotesize
\begin{tabular}{cccccccc}
\hline
Globular & Dist. & $S_{min}$(1400) &  $V_{\rm esc}$ & Metallicity & Mass & Number of & Search \\
Cluster & (kpc) & ($\mu$Jy) & (km~s$^{-1}$) & log[Fe/H] & $\msun$ & Young Pulsars & Reference \\
\hline
NGC~6540 & 5.3 & 77.5 & 27.6 & $-$1.35 & 36,400 & 0 & Possenti unpublished 2010  \\ 
NGC~6544 & 3.0 & 38.8 & 93.5 & $-$1.40 & 108,000 & 0 & Possenti unpublished 2010  \\ 
NGC~6541 & 7.5 & 9.6 & 42.2 & $-$1.81 & 572,000 & 0 & Lynch unpublished 2010  \\ 
NGC~6558 & 7.4 & 63.3 & 32.1 & $-$1.32 & 98,400 & 0 & Possenti unpublished 2010  \\ 
NGC~6584 & 13.5 & 54.8 & 24.3 & $-$1.50 & 303,000 & 0 & Possenti unpublished 2010  \\ 
NGC~6624 & 7.9 & 16.0 & 35.3 & $-$0.44 & 257,000 & 2 & Lynch et al 2011  \\
M~28 & 5.5 & 4.1 & 63.8 & $-$1.32 & 551,000 & 0 & Lynch unpublished 2010  \\ 
NGC~6642 & 8.1 & 54.8 & 30.7 & $-$1.26 & 109,000 & 0 & Possenti unpublished 2010  \\ 
NGC~6652 & 10.0 & 77.5 & 37.5 & $-$0.81 & 109,000 & 0 & Possenti unpublished 2010  \\ 
M~22 & 3.2 & 5.7 & 44.7 & $-$1.70 & 644,000 & 0 & Lynch et al 2011  \\ 
NGC~6681 & 9.0 & 77.5 & 39.3 & $-$1.62 & 179,000 & 0 & Possenti unpublished 2010  \\ 
NGC~6712 & 6.9 & 6.7 & 27.7 & $-$1.02 & 257,000 & 0 & Lynch et al 2011  \\ 
NGC~6717 & 7.1 & 54.8 & 21.6 & $-$1.26 & 47,500 & 0 & Possenti unpublished 2010  \\ 
NGC~6723 & 8.7 & 54.8 & 27.3 & $-$1.10 & 357,000 & 0 & Possenti unpublished 2010  \\ 
NGC~6749 & 7.9 & 32 & 20.4 & $-$1.60 & 123,000 & 0 & Hessels et al 2007  \\
NGC~6752 & 4.0 & 31.5 & 32.9 & $-$1.54 & 317,000 & 0 & Possenti unpublished 2010  \\ 
NGC~6760 & 7.4 & 37 & 40.1 & $-$0.40 & 357,000 & 0 & Hessels et al 2007  \\ 
NGC~6779 & 9.4 & 25 & 28.7 & $-$1.98 & 230,000 & 0 & Hessels et al 2007  \\ 
Pal~10 & 5.9 & 23 & 17.2 & $-$0.10 & 53,100 & 0 & Hessels et al 2007  \\
NGC~6809 & 5.4 & 77.5 & 19.7 & $-$1.94 & 269,000 & 0 & Possenti unpublished 2010  \\ 
NGC~6838 & 4.0 & 19 & 16.7 & $-$0.78 & 43,000 & 0 & Hessels et al 2007  \\ 
NGC~6934 & 15.6 & 28 & 28.1 & $-$1.47 & 295,000 & 0 & Hessels et al 2007  \\ 
NGC~6981 & 17.0 & 6.0 & 16.6 & $-$1.42 & 168,000 & 0 & Lynch \& Ransom 2011  \\ 
NGC~7006 & 41.2 & 19 & 19.8 & $-$1.52 & 303,000 & 0 & Hessels et al 2007  \\ 
NGC~7078 & 10.4 & 21 & 62.1 & $-$2.37 & 1,190,000 & 0 & Hessels et al 2007  \\ 
NGC~7089 & 11.5 & 6.0 & 48.1 & $-$1.65 & 104,000 & 0 & Lynch \& Ransom 2011  \\ 
NGC~7099 & 8.1 & 10.8 & 34.1 & $-$2.27 & 241,000 & 0 & Ransom et al 2004  \\ 
Pal~12 & 19.0 & 6.2 & 5.5 & $-$0.85 & 15,900 & 0 & Lynch \& Ransom 2011  \\ 
Pal~13 & 26.0 & 21 & 3.5 & $-$1.88 & 6,500 & 0 & Hessels et al 2007  \\ 

\hline
\end{tabular}
\smallskip
\end{center}
\end{table*}

\begin{table*}[h]
\begin{center}
\caption{Binomial analysis of young pulsars in globular clusters.} 
\smallskip
\tiny
\begin{tabular}{cccccccccccc}
\hline

Globular & Median & 2.5$^{\rm th}$ & 97.5$^{\rm th}$ & log($f_{R1}$) & log($f_{R2}$) & log($f_{R3}$) & log(${\cal R}$)$f_{R1}$ & log(${\cal R}$)$f_{R2}$ & log(${\cal R}$)$f_{R3}$ & $\rm N_{predicted}$ \\
Cluster & & percentile & percentile & 50 km~s$^{-1}$ & 20 km~s$^{-1}$ & 10 km~s$^{-1}$ & (psrs cen$^{-1}$) & (psrs cen$^{-1}$) & (psrs cen$^{-1}$) & mass model \\
\hline
47~Tuc & 16 & 0 & 88 & $-$0.39 & 0.00 & 0.00 & $-$3.04 & $-$3.43 & $-$3.43 & 1 \\
NGC~288 & 3 & 0 & 17 & $-$2.30 & $-$1.16 & $-$0.42 & $-$1.85 & $-$3.00 & $-$3.74 & 0 \\
NGC~1261 & 53 & 1 & 297 & $-$4.08 & $-$2.88 & $-$1.99 & 1.15 & $-$0.03 & $-$0.92 & 0 \\
Pal~2 & 167 & 6 & 968 & $-$1.37 & $-$0.38 & $-$0.02 & $-$1.03 & $-$2.03 & $-$2.40 & 0 \\
NGC~1851 & 21 & 0 & 115 & $-$0.66 & $-$0.03 & 0.00 & $-$2.65 & $-$3.28 & $-$3.32 & 0 \\
NGC~2298 & 3 & 0 & 16 & $-$1.89 & $-$0.79 & $-$0.17 & $-$2.26 & $-$3.37 & $-$3.99 & 3 \\
NGC~2808 & 24 & 0 & 133 & $-$0.34 & 0.00 & 0.00 & $-$2.92 & $-$3.26 & $-$3.26 & 0 \\
E3 & 17 & 0 & 93 & $-$4.24 & $-$3.05 & $-$2.15 & 0.82 & $-$0.36 & $-$1.25 & 0 \\
NGC~3201 & 6 & 0 & 36 & $-$1.66 & $-$0.60 & $-$0.08 & $-$2.19 & $-$3.26 & $-$3.77 & 0 \\
NGC~4147 & 34 & 1 & 192 & $-$1.90 & $-$0.79 & $-$0.17 & $-$1.20 & $-$2.31 & $-$2.93 & 0 \\
NGC~4372 & 9 & 0 & 49 & $-$1.70 & $-$0.62 & $-$0.09 & $-$1.99 & $-$3.06 & $-$3.59 & 0 \\
NGC~4590 & 28 & 1 & 155 & $-$1.90 & $-$0.80 & $-$0.18 & $-$1.28 & $-$2.39 & $-$3.01 & 0 \\
NGC~4833 & 16 & 0 & 87 & $-$1.21 & $-$0.27 & 0.00 & $-$2.22 & $-$3.16 & $-$3.43 & 0 \\
NGC~5024 & 29 & 1 & 163 & $-$1.15 & $-$0.23 & 0.00 & $-$2.02 & $-$2.94 & $-$3.17 & 0 \\
NGC~5053 & 29 & 1 & 162 & $-$2.83 & $-$1.65 & $-$0.82 & $-$0.35 & $-$1.52 & $-$2.35 & 0 \\
NGC~5139 & 6 & 0 & 36 & $-$0.50 & 0.00 & 0.00 & $-$3.35 & $-$3.85 & $-$3.86 & 8 \\
NGC~5272 & 10 & 0 & 57 & $-$1.03 & $-$0.16 & 0.00 & $-$2.61 & $-$3.47 & $-$3.64 & 1 \\
NGC~5286 & 37 & 1 & 205 & $-$0.64 & $-$0.03 & 0.00 & $-$2.42 & $-$3.04 & $-$3.07 & 0 \\
NGC~5466 & 27 & 0 & 149 & $-$2.74 & $-$1.57 & $-$0.75 & $-$0.46 & $-$1.63 & $-$2.45 & 0 \\
Pal~5 & 101 & 3 & 575 & $-$4.16 & $-$2.96 & $-$2.07 & 1.51 & 0.32 & $-$0.56 & 0 \\
NGC~5897 & 4 & 0 & 25 & $-$2.30 & $-$1.15 & $-$0.41 & $-$1.73 & $-$2.88 & $-$3.62 & 0 \\
NGC~5904 & 7 & 0 & 39 & $-$0.75 & $-$0.05 & 0.00 & $-$3.04 & $-$3.73 & $-$3.80 & 2 \\
NGC~5927 & 15 & 0 & 84 & $-$1.13 & $-$0.23 & 0.00 & $-$2.32 & $-$3.23 & $-$3.46 & 0 \\
NGC~5946 & 29 & 1 & 164 & $-$1.49 & $-$0.46 & $-$0.03 & $-$1.68 & $-$2.71 & $-$3.13 & 0 \\
NGC~5986 & 2 & 0 & 14 & $-$1.03 & $-$0.17 & 0.00 & $-$3.30 & $-$4.16 & $-$4.34 & 3 \\
M~80 & 3 & 0 & 19 & $-$0.73 & $-$0.05 & 0.00 & $-$3.43 & $-$4.11 & $-$4.16 & 6 \\
NGC~6121 & 3 & 0 & 20 & $-$1.13 & $-$0.22 & 0.00 & $-$3.03 & $-$3.94 & $-$4.16 & 0 \\
ESO452 & 17 & 0 & 98 & $-$3.36 & $-$2.17 & $-$1.30 & $-$0.04 & $-$1.23 & $-$2.10 & 0 \\
NGC~6144 & 20 & 0 & 113 & $-$2.23 & $-$1.09 & $-$0.36 & $-$1.10 & $-$2.24 & $-$2.97 & 0 \\
NGC~6139 & 39 & 1 & 218 & $-$0.53 & $-$0.01 & 0.00 & $-$2.51 & $-$3.03 & $-$3.05 & 0 \\
NGC~6171 & 10 & 0 & 59 & $-$1.50 & $-$0.47 & $-$0.04 & $-$2.13 & $-$3.16 & $-$3.60 & 0 \\
NGC~6205 & 7 & 0 & 37 & $-$0.97 & $-$0.14 & 0.00 & $-$2.82 & $-$3.65 & $-$3.80 & 1 \\
NGC~6218 & 6 & 0 & 35 & $-$1.34 & $-$0.36 & $-$0.01 & $-$2.51 & $-$3.50 & $-$3.84 & 0 \\
NGC~6235 & 35 & 1 & 197 & $-$2.01 & $-$0.89 & $-$0.23 & $-$1.08 & $-$2.20 & $-$2.86 & 0 \\
NGC~6254 & 6 & 0 & 36 & $-$1.30 & $-$0.33 & $-$0.01 & $-$2.55 & $-$3.53 & $-$3.85 & 0 \\
Pal~15 & 878 & 30 & 5409 & $-$3.77 & $-$2.58 & $-$1.70 & 2.07 & 0.88 & 0.00 & 0 \\
NGC~6266 & 5 & 0 & 29 & $-$0.13 & 0.00 & 0.00 & $-$3.80 & $-$3.94 & $-$3.94 & 3 \\
NGC~6273 & 20 & 0 & 110 & $-$0.54 & $-$0.01 & 0.00 & $-$2.80 & $-$3.32 & $-$3.34 & 0 \\
NGC~6284 & 69 & 2 & 391 & $-$1.34 & $-$0.35 & $-$0.01 & $-$1.46 & $-$2.44 & $-$2.78 & 0 \\
NGC~6287 & 33 & 1 & 185 & $-$1.26 & $-$0.30 & 0.00 & $-$1.85 & $-$2.81 & $-$3.11 & 0 \\
NGC~6293 & 23 & 0 & 130 & $-$0.90 & $-$0.10 & 0.00 & $-$2.38 & $-$3.16 & $-$3.28 & 0 \\
NGC~6304 & 9 & 0 & 51 & $-$0.99 & $-$0.15 & 0.00 & $-$2.69 & $-$3.53 & $-$3.69 & 0 \\
M~92 & 2 & 0 & 13 & $-$0.76 & $-$0.06 & 0.00 & $-$3.57 & $-$4.28 & $-$4.34 & 2 \\
NGC~6325 & 15 & 0 & 86 & $-$0.87 & $-$0.10 & 0.00 & $-$2.58 & $-$3.36 & $-$3.46 & 0 \\
NGC~6333 & 16 & 0 & 88 & $-$1.01 & $-$0.16 & 0.00 & $-$2.42 & $-$3.27 & $-$3.43 & 0 \\
NGC~6342 & 288 & 40 & 1047 & $-$1.61 & $-$0.56 & $-$0.06 & $-$0.56 & $-$1.62 & $-$2.11 & 0 \\
NGC~6355 & 31 & 1 & 176 & $-$0.93 & $-$0.12 & 0.00 & $-$2.21 & $-$3.02 & $-$3.15 & 0 \\
Liller~1 & 3 & 0 & 21 & $-$0.91 & $-$0.11 & 0.00 & $-$3.25 & $-$4.05 & $-$4.16 & 0 \\
Ter~1 & 16 & 0 & 90 & $-$3.06 & $-$1.88 & $-$1.03 & $-$0.36 & $-$1.55 & $-$2.40 & 0 \\
NGC~6388 & 25 & 0 & 142 & $-$0.04 & 0.00 & 0.00 & $-$3.20 & $-$3.24 & $-$3.24 & 1 \\
NGC~6402 & 22 & 0 & 124 & $-$0.97 & $-$0.14 & 0.00 & $-$2.32 & $-$3.16 & $-$3.30 & 0 \\
NGC~6401 & 43 & 1 & 244 & $-$0.99 & $-$0.15 & 0.00 & $-$2.01 & $-$2.85 & $-$3.00 & 0 \\
NGC~6397 & 1 & 0 & 8 & $-$0.73 & $-$0.05 & 0.00 & $-$3.90 & $-$4.59 & $-$4.64 & 1 \\
Pal~6 & 1 & 0 & 10 & $-$1.36 & $-$0.37 & $-$0.02 & $-$3.27 & $-$4.26 & $-$4.62 & 2 \\
NGC~6426 & 55 & 1 & 308 & $-$2.18 & $-$1.04 & $-$0.33 & $-$0.72 & $-$1.85 & $-$2.56 & 0 \\
Ter~5 & 3 & 0 & 18 & $-$0.69 & $-$0.03 & 0.00 & $-$3.47 & $-$4.12 & $-$4.16 & 2 \\
NGC~6440 & 10 & 2 & 34 & $-$0.22 & 0.00 & 0.00 & $-$3.41 & $-$3.64 & $-$3.64 & 3 \\
NGC~6441 & 8 & 0 & 45 & $-$0.12 & 0.00 & 0.00 & $-$3.62 & $-$3.74 & $-$3.74 & 3 \\
Ter~6 & 2 & 0 & 13 & $-$0.99 & $-$0.15 & 0.00 & $-$3.34 & $-$4.19 & $-$4.34 & 1 \\
NGC~6453 & 56 & 2 & 316 & $-$1.64 & $-$0.58 & $-$0.07 & $-$1.24 & $-$2.30 & $-$2.81 & 0 \\
UKS~1 & 15 & 0 & 86 & $-$1.49 & $-$0.46 & $-$0.03 & $-$1.97 & $-$3.00 & $-$3.42 & 0 \\
NGC~6496 & 34 & 1 & 189 & $-$1.80 & $-$0.71 & $-$0.13 & $-$1.30 & $-$2.39 & $-$2.97 & 0 \\
Ter~9 & 4 & 0 & 25 & $-$2.70 & $-$1.53 & $-$0.72 & $-$1.33 & $-$2.50 & $-$3.32 & 0 \\
NGC~6517 & 2 & 0 & 13 & $-$0.24 & 0.00 & 0.00 & $-$4.10 & $-$4.34 & $-$4.34 & 3 \\
NGC~6522 & 15 & 0 & 84 & $-$0.88 & $-$0.10 & 0.00 & $-$2.58 & $-$3.36 & $-$3.46 & 0 \\
NGC~6528 & 16 & 0 & 88 & $-$1.44 & $-$0.42 & $-$0.03 & $-$1.99 & $-$3.01 & $-$3.40 & 0 \\
NGC~6535 & 9 & 0 & 50 & $-$2.67 & $-$1.50 & $-$0.70 & $-$1.01 & $-$2.17 & $-$2.98 & 0 \\
NGC~6539 & 13 & 0 & 74 & $-$1.07 & $-$0.19 & 0.00 & $-$2.45 & $-$3.33 & $-$3.53 & 0 \\
NGC~6540 & 10 & 0 & 57 & $-$1.38 & $-$0.38 & $-$0.02 & $-$2.25 & $-$3.25 & $-$3.62 & 0 \\
NGC~6544 & 2 & 0 & 13 & $-$0.16 & 0.00 & 0.00 & $-$4.17 & $-$4.34 & $-$4.34 & 0 \\
NGC~6541 & 3 & 0 & 18 & $-$0.88 & $-$0.10 & 0.00 & $-$3.28 & $-$4.06 & $-$4.16 & 3 \\
NGC~6558 & 16 & 0 & 90 & $-$1.20 & $-$0.26 & 0.00 & $-$2.23 & $-$3.16 & $-$3.43 & 0 \\
NGC~6584 & 51 & 1 & 287 & $-$1.54 & $-$0.50 & $-$0.05 & $-$1.39 & $-$2.43 & $-$2.88 & 0 \\
NGC~6624 & 21 & 5 & 58 & $-$1.09 & $-$0.20 & 0.00 & $-$2.23 & $-$3.11 & $-$3.32 & 0 \\
M~28 & 1 & 0 & 7 & $-$0.45 & 0.00 & 0.00 & $-$4.18 & $-$4.63 & $-$4.64 & 10 \\
NGC~6642 & 17 & 0 & 93 & $-$1.25 & $-$0.30 & 0.00 & $-$2.15 & $-$3.11 & $-$3.40 & 0 \\
NGC~6652 & 38 & 1 & 213 & $-$1.02 & $-$0.16 & 0.00 & $-$2.04 & $-$2.90 & $-$3.06 & 0 \\
M~22 & 0 & 0 & 4 & $-$0.82 & $-$0.07 & 0.00 & $-$1.74$^a$ & $-$2.66$^a$ & $-$4.64$^a$ & 0 \\
NGC~6681 & 30 & 1 & 168 & $-$0.96 & $-$0.13 & 0.00 & $-$2.20 & $-$3.02 & $-$3.16 & 0 \\
NGC~6712 & 2 & 0 & 12 & $-$1.38 & $-$0.38 & $-$0.02 & $-$2.96 & $-$3.95 & $-$4.32 & 1 \\
NGC~6717 & 13 & 0 & 72 & $-$1.69 & $-$0.62 & $-$0.09 & $-$1.83 & $-$2.90 & $-$3.43 & 0 \\
NGC~6723 & 19 & 0 & 108 & $-$1.40 & $-$0.39 & $-$0.02 & $-$1.96 & $-$2.96 & $-$3.33 & 0 \\
NGC~6749 & 9 & 0 & 53 & $-$1.76 & $-$0.67 & $-$0.12 & $-$1.92 & $-$3.00 & $-$3.56 & 0 \\
NGC~6752 & 3 & 0 & 17 & $-$1.17 & $-$0.24 & 0.00 & $-$2.99 & $-$3.91 & $-$4.16 & 1 \\
NGC~6760 & 9 & 0 & 53 & $-$0.94 & $-$0.13 & 0.00 & $-$2.74 & $-$3.56 & $-$3.69 & 0 \\
NGC~6779 & 10 & 0 & 58 & $-$1.33 & $-$0.35 & $-$0.01 & $-$2.30 & $-$3.28 & $-$3.62 & 0 \\
Pal~10 & 4 & 0 & 24 & $-$1.98 & $-$0.86 & $-$0.21 & $-$2.06 & $-$3.17 & $-$3.82 & 0 \\

\end{tabular}
\smallskip
\label{table:gcbin}
\end{center}
\end{table*}

\addtocounter{table}{-1}
\begin{table*}[h]
\begin{center}
\caption{Continued}
\smallskip
\tiny
\begin{tabular}{cccccccccccc}
\hline

Globular  & Median & 2.5$^{\rm th}$ & 97.5$^{\rm th}$ & log($f_{R1}$) & log($f_{R2}$) & log($f_{R3}$) & log(${\cal R}$)$f_{R1}$ & log(${\cal R}$)$f_{R2}$ & log(${\cal R}$)$f_{R3}$ & $\rm N_{predicted}$ \\
Cluster  & & percentile & percentile & 50 km~s$^{-1}$ & 20 km~s$^{-1}$ & 10 km~s$^{-1}$ & (psrs cen$^{-1}$) & (psrs cen$^{-1}$) & (psrs cen$^{-1}$) & mass model \\
\hline
NGC~6809 & 10 & 0 & 59 & $-$1.80 & $-$0.71 & $-$0.13 & $-$1.83 & $-$2.92 & $-$3.50 & 0 \\
NGC~6838 & 2 & 0 & 12 & $-$2.01 & $-$0.90 & $-$0.23 & $-$2.32 & $-$3.44 & $-$4.10 & 0 \\
NGC~6934 & 33 & 1 & 184 & $-$1.36 & $-$0.37 & $-$0.01 & $-$1.75 & $-$2.75 & $-$3.10 & 0 \\
NGC~6981 & 8 & 0 & 46 & $-$2.02 & $-$0.90 & $-$0.24 & $-$1.71 & $-$2.83 & $-$3.49 & 0 \\
NGC~7006 & 237 & 8 & 1384 & $-$1.80 & $-$0.71 & $-$0.13 & $-$0.46 & $-$1.55 & $-$2.13 & 0 \\
NGC~7078 & 11 & 0 & 60 & $-$0.48 & 0.00 & 0.00 & $-$3.11 & $-$3.59 & $-$3.60 & 2 \\
NGC~7089 & 4 & 0 & 24 & $-$0.74 & $-$0.05 & 0.00 & $-$3.30 & $-$3.98 & $-$4.04 & 3 \\
NGC~7099 & 4 & 0 & 22 & $-$1.13 & $-$0.22 & 0.00 & $-$2.90 & $-$3.81 & $-$4.03 & 1 \\
Pal~12 & 10 & 0 & 59 & $-$3.45 & $-$2.26 & $-$1.39 & $-$0.18 & $-$1.37 & $-$2.25 & 0 \\
Pal~13 & 79 & 2 & 446 & $-$4.04 & $-$2.85 & $-$1.96 & 1.29 & 0.10 & $-$0.78 & 0 \\
\hline 
\multicolumn{11}{l}{\small $^a$ Values are upper limits assuming M22 contains one pulsar.} \\
\multicolumn{11}{l}{\small $f_R$ is the retention fraction for a given velocity dispersion.} \\
\multicolumn{11}{l}{\small ${\cal R}$ is the birthrate in pulsars per century for the median value.} \\
\multicolumn{11}{l}{\small Birthrates are upper limits for a given globular cluster except} \\
\multicolumn{11}{l}{\small for those with detected young pulsars. } \\
\end{tabular} 
\smallskip
\end{center}
\end{table*}

\end{document}